\documentclass[pra,superscriptaddress,reprint,longbibliography, floatfix]{revtex4-1}
\usepackage{graphicx}
\usepackage{url}
\begin{document}

\title{Creating a Physicist: The Impact of Informal Physics Programs on University Student Development}

\date{\today}

\author{Callie Rethman}
\affiliation{Department of Physics and Astronomy, Texas A\&M University,
		College Station, TX~~77845}
\email[send correspondence to: ]{etanya@tamu.edu}
\author{Jonathan Perry}
\affiliation{Department of Physics, University of Texas,
		Austin, TX~~78712}
\author{Jonan Donaldson}
\affiliation{Center for Teaching Excellence, Texas A\&M University, College Station, TX ~~77845}
\author{Daniel Choi}
\affiliation{Center for Teaching Excellence, Texas A\&M University, College Station, TX ~~77845}
\author{Tatiana Erukhimova}
\affiliation{Department of Physics and Astronomy, Texas A\&M University,
		College Station, TX~~77845}

\begin{abstract}
Physics outreach programs provide a critical context for informal experiences that promote the transition from new student to contributing physicist. Prior studies have suggested a positive link between participation in informal physics outreach programs and the development of a student's physics identity. In this study, we adopt a student-focused investigation to explore the effects of informal programs on dimensions of physics identity, sense of community, 21st century skill development, and motivation. We employed a mixed methods study combining a survey instrument (117 responses) and interviews (35) with current and former undergraduate and graduate students who participated in five programs through a physics and astronomy department at a large land-grant university. To examine interviews, we employed a framework based on situated learning theory, transformative learning theory, and the Dynamic Systems Model of Role Identity. Our findings, based on self-reported data, show that students who facilitated informal physics programs positively developed their physics identity, experienced increased sense of belonging to the physics community, and developed 21st century career skills. Specifically, students reported positive benefits to their communication, teamwork and networking, and design skills. The benefits of these programs can be achieved by departments of any size without significant commitment of funds or changes to curriculum. 

\end{abstract}

\maketitle

\section{Introduction}

Many physics departments and national labs in the United States run informal programs. These programs are often called “public outreach” \cite{Graur2018}, reflecting a historical understanding of their main purpose: building a bridge between “ivory tower” physicists and the general public, as well as providing unique opportunities for engaging children in STEM, especially unprivileged children \cite{Leshner2007, Holt2015, European2002, Nature2017, NAP2015, NAP2009}. Most American scientists agree that they should “take an active role in public policy debates about issues related to science and technology” \cite{Rainie2015}. There is an ongoing discussion on how to be more effective in communicating scientific advances to various audiences \cite{Nature2017, Besley2018}. There have been calls from prominent scientists to train future generations of scientists to be effective science communicators \cite{Leshner2007, Brownell2013} and to recognize public outreach effort as an integral part of scientists' careers in academia \cite{Graur2018, Foster2010}. 

Prior literature exhibits a consensus on the positive impact of out-of-school programs on children. These programs increase childrens' understanding and interest in STEM and generate enthusiasm for science \cite{NAP2015, NAP2009, Bartley2009, Wulf2012}. They can be especially impactful for children from underserved communities and females who otherwise may not be interested in science simply due to their lack of exposure to science programs and role models \cite{Dabney2012, Schibeci2006}. Major professional organizations of physicists, such as the American Physical Society \cite{APS}, American Association of Physics Teachers \cite{AAPT}, the Optical Society \cite{OSA}, the International Society for Optics and Photonics \cite{SPIE}, as well as the Society of Physics Students \cite{SPS}, make considerable efforts to share outreach resources that help their members engage with the general public. Funding agencies, such as the National Science Foundation, encourage engagement between researchers and various communities through outreach as part of their ``broader impact" requirements \cite{NSF}. Undergraduate and graduate students have served as facilitators of physics outreach programs for decades, and many program organizers would consider it obvious that these students benefit from participation in the outreach programs. In this paper we use the terms ``informal" and ``outreach" interchangeably to describe programs in which students participate that are outside formal curriculum.

There has been increasing interest in understanding how undergraduate and graduate student facilitation of these programs supports the development of a physics and STEM identity, enhances retention and persistence, and supports a feeling of community \cite{Finkelstein2008, Hinko2012, Hinko2014, Hinko2016, Fracchiolla2016, Fracchiolla2018, Prefontaine2018, Fracchiolla2020, Sauncy2015, STEM2015, Quan2019}. The results of this paper add to the growing understanding of how informal physics programs provide a platform for broader interactions between an individual student and the STEM community and equip university students with the skills needed for the 21st century careers \cite{Heron2016}. 

The perspective that informal physics programs are beneficial only for the public is detrimental, as it places them outside of university research and teaching missions \cite{Hinko2012} and makes them low priority for institutional support. Studies reporting on successful programs reflect on university-community partnership, effective communication, passion for science, reducing stereotypes, and longevity \cite{Finkelstein2008, Bartley2009, Wulf2012, Miranda2018, Hayes2020, Greenler1993}. These programs are usually viewed as part of a service mission of a physics department or as a recruitment tool \cite{Hinko2012, Steward2013}, though some university professors may dissuade their students from engaging in outreach believing their time is better spent on research \cite{Graur2018}. 

There is not much literature on the impact of participation in informal physics programs on university students. In 2008, Finkelstein and Mayhew presented the results of a university-community partnership, Partnerships for Informal Science Education in the Community (PISEC) at the University of Colorado Boulder, where university students were mentored to teach youth in an after-school community setting \cite{Finkelstein2008}. In a subsequent study, Hinko and Finkelstein \cite{Hinko2012} reported that university students had positive shifts in their perspectives of teaching and learning, and improved their science communication skills through participating in PISEC. They encouraged a shift from “outreach” to “partnership”, emphasizing a win-win situation for both universities and communities. Through further exploration of PISEC, Hinko et al. \cite{Hinko2014} constructed a framework for the assessment of scientific language for physics students explaining (informal teaching) concepts to non-expert audiences. 

Teaching experience is a crucial aspect of formal physics training \cite{Feldon2011, Drane2014}. Many graduate students and some undergraduate students acquire teaching experience through teaching assistantships \cite{Otero2010}, but these are formal roles that are constrained by the curriculum. Hinko et al. \cite{Hinko2016} argued that an overlooked area of the physics teaching experience for undergraduate and graduate physics students is informal physics programs where these students serve as facilitators. As compared to formal teaching assistantships, informal physics programs provide less constraints, more ownership, more room for initiative, more flexibility in time commitment, and more excitement. This may translate into richer teaching opportunities, formal and informal, for students.

Prior work indicates that development of a physics identity could help students choose physics as a career and persist in the field \cite{Hazari2010}. Discipline-based identity, intertwined with the development of motivational beliefs, increased self-efficacy, sense of belonging, external recognition, and “real-world” experience could be the leading factors in students’ persistence in, or attrition from, physics and other STEM fields; thus, it has a potential for enhanced retention among students, especially among underrepresented minority populations \cite{Perez2014, Sauncy2015, STEM2015, Quan2019, Sawtelle2012, Zwolak2017, Thiry2011, Kalender2019, Hyater-Adams2018, Hyater-Adams2019, Lewis2016, Hazari2013a, Hazari2013b, Close2016, Irving2015}. Recent work from Fracchiolla et al. applied a community of practice framework to study one after school program focusing on aspects including connections within the physics community, sense of belonging, and development of physics identity \cite{Fracchiolla2020}. Their work suggests that volunteering in informal physics programs could have a positive influence on the growth of a university student's physics identity. 

Informal physics programs differ in terms of their target audience, facilitators, settings, modes of implementation, scale, frequency, longevity, and institutional support \cite{Fracchiolla2018}. Prior studies examined a relatively small number of participants drawn from a limited range of informal physics programs. In this paper, we present the findings from a mixed methods study on the impact of different kinds of informal physics programs on a large number of undergraduate and graduate students facilitating these programs at Texas A\&M University. The Department of Physics \& Astronomy at Texas A\&M runs several nationally-recognized informal STEM learning programs. They span a wide range of activities and public audiences -- from the \textit{Texas A\&M Physics \& Engineering Festival} where people can spend all day playing with hands-on demonstrations, talking to and learning from the top-notch researchers and astronauts -- to the \textit{Just Add Science \& Game Day Physics} programs which bring the excitement of physics to places where people already are, such as heritage festivals, football games, or community festivals. The \textit{Physics Show} targets organized groups of K-12 students on campus. In the \textit{Real Physics Live} program, university students create entertaining educational videos which illustrate important physics concepts using demonstration experiments.

A goal of these programs is to make science exciting, understandable, and accessible to the general public. Another equally important goal is to provide opportunities for undergraduate and graduate students’ personal, academic, and professional growth. One program, \textit{Discover, Explore, and Enjoy Physics \& Engineering} (DEEP) was designed to be student-centered, with the main focus on the experience of the university students, while other programs described in this paper evolved over time from being considered as “public outreach" to becoming an integral part of university students' educational experience.

We conducted a student-focused investigation examining the impact of informal physics programs on undergraduate and graduate student volunteers helping to run these programs. We explored the effects of Texas A\&M informal physics programs on
(a) establishing a student’s identity as a physicist and a STEM professional; (b) students’ sense of belonging to the physics community and the broader STEM community; (c) students’ development of soft skills, such as communication, teamwork, design, and conceptual understanding; and (d) students’ experiences, such as seeing new perspectives, motivation, interest development, and empowerment. Also emergent in our analyses were unique impacts of outreach programs on female students who are traditionally underrepresented in physics. The results of this study could be of potential interest to every physics department and physics educator, since some of the informal physics programs do not require a large budget or any changes in the curriculum. 

\section{Program Structure}
\begin{figure}[h]
\centering
\includegraphics[width=7.6cm]{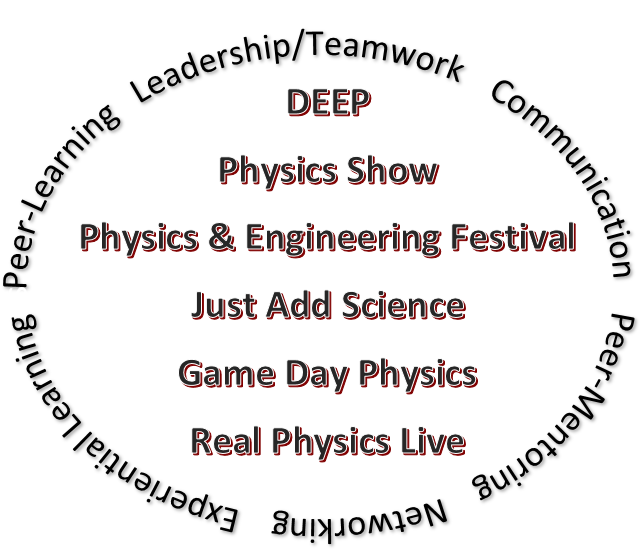}
\caption{\label{Programs} Schematic of the program principles and their associated informal physics programs at Texas A\&M. }
\end{figure}

We analyzed the impact of five informal programs run by the Department of Physics \& Astronomy at Texas A\&M on the university students who facilitate these programs. Table \ref{PhysicsOutreach} lists the years of implementation of these programs as well as an approximate number of students participating every year. Although each of these programs were founded at different times and with a different target audience in mind, they all provide university students with potential opportunities for leadership and teamwork, experiential learning, peer-mentoring and peer-learning, networking within the different populaces of an academic department, and developing important communication skills (Figure \ref{Programs}). We describe the program Discover, Explore, and Enjoy Physics \& Engineering (DEEP) in more detail since this program was designed with a focus on enriching university students learning and experience.

\textbf{\emph{Discover, Explore and Enjoy Physics \& Engineering}} (DEEP) is a hands-on, peer-learning community. On average, there are 60 undergraduate students and 13 graduate students (DEEP Mentors) who participate in the program each year (2012-2021). Students work throughout the academic year in teams of 5-10, side-by-side with their peers and graduate student mentors on research, concept, design, and fabrication of physics demonstration experiments. Though most students come from science and engineering majors, participation is open to students from any discipline. The same student teams present their experiments through other informal physics programs described in this section. The demonstrations fabricated by students are added to the pool of demonstration experiments available for all physics and astronomy courses.

\begin{table}[h]
\caption{\label{PhysicsOutreach} Approximate number of annual student participants and the inaugural year for the Texas A\&M informal physics programs included in this study.}
\begin{ruledtabular}
\begin{tabular}{ccc }
 Program title &  Initial year & Number  \\
  & & of students\\
 \hline
 DEEP & 2012 & 70 \\
 Physics \& Engineering Festival & 2003 & 300-400 \\
 Physics Show & 2007 & 200 \\
 Real Physics Live & 2016 & 11 \\
 Just Add Science/Game Day Physics & 2015 & 70\\
\end{tabular}
\end{ruledtabular}
\end{table}

This program was designed with the intention that through these collaborative hands-on extracurricular activities, students learn physics concepts more deeply, get more opportunities for interactions with peers and professors outside the classroom, develop collaboration skills through team interactions, and increase communication skills as a result of presentations to a wide range of audiences. The core goal of this program is to deepen students’ physics content knowledge through transferable skills (i.e., teamwork, communication ability, and ethics) and hands-on experiences, utilizing each individual’s science background and identity to enhance their STEM learning experience through peer learning communities aimed at small group and individualized instruction.

The DEEP program facilitates peer mentoring which includes not only undergraduate students interacting with each other across all classifications but also graduate students mentoring undergraduates in their group. The latter is fairly unique as graduate and undergraduate student populations usually do not interact outside of formal settings.


Demonstration experiments cover a broad variety of topics from physics, chemistry, electrical and computer engineering, etc. Students are encouraged to be creative with ideas for demonstrations. They also prepare a poster and a narrative explaining the underlying concepts at a level accessible for visitors of all educational levels. Undergraduate students are involved in every aspect of design, fabrication, and presentation of the experiments, gaining invaluable experience. Students often enter the program as freshmen and develop these skills over the course of their undergraduate careers. Graduate students leading teams of undergraduates are provided with opportunities to acquire leadership and mentoring experience: they build a collaborative research team and lead this team through research, fabrication, and presentation of their projects. 

\begin{figure}[h]
\centering
\includegraphics[width=8.6cm]{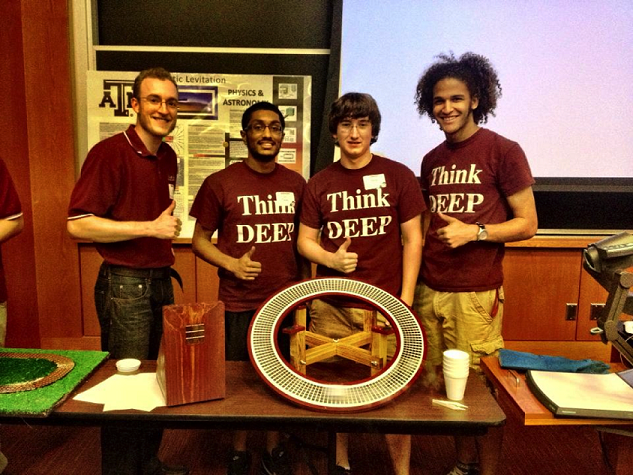}
\caption{\label{super} DEEP team members with
their superconducting train track. This demonstration experiment was part of Nobel Laureate David Lee's public talk at the 2013 Physics \& Engineering Festival. Photo by Natasha Sheffield.}
\end{figure}

One example of a DEEP demonstration experiment designed and fabricated mostly by freshman students is the superconducting train on a magnetic track (Figure ~\ref{super}). This single experiment teaches nearly all of the basic concepts of electricity and magnetism and several key concepts from advanced physics courses. It connects students with one of the most advanced and fascinating transportation technologies: magnetic levitation trains that are currently being tested in Japan, Europe, and the USA. Finally, it is mesmerizing for anyone, from small kids to adults, to watch how the train levitates while going round and round the magnetic track as if being held by some mysterious force. It is therefore not surprising that this demonstration experiment is a huge hit at Physics Shows, a favorite of the public at the Physics \& Engineering Festivals, and is regularly shown in the classroom. 

A good DEEP demonstration experiment does not have to be technically advanced or expensive to be of high educational value. Another example conceived, designed, and fabricated by DEEP students is a simple lever demonstration. Although simple, it teaches a number of important concepts of mechanics and provides a vivid explanation of the operation of construction cranes and lifting machines. Another exciting demonstration built by DEEP students, Methane Bubbles, was featured on the front page of the SPS Observer \cite{SPS2014} as seen in Figure 3. Displaying this demonstration experiment requires team discipline and following strict safety rules. 

\begin{figure}[h]
\centering
\includegraphics[width=6cm]{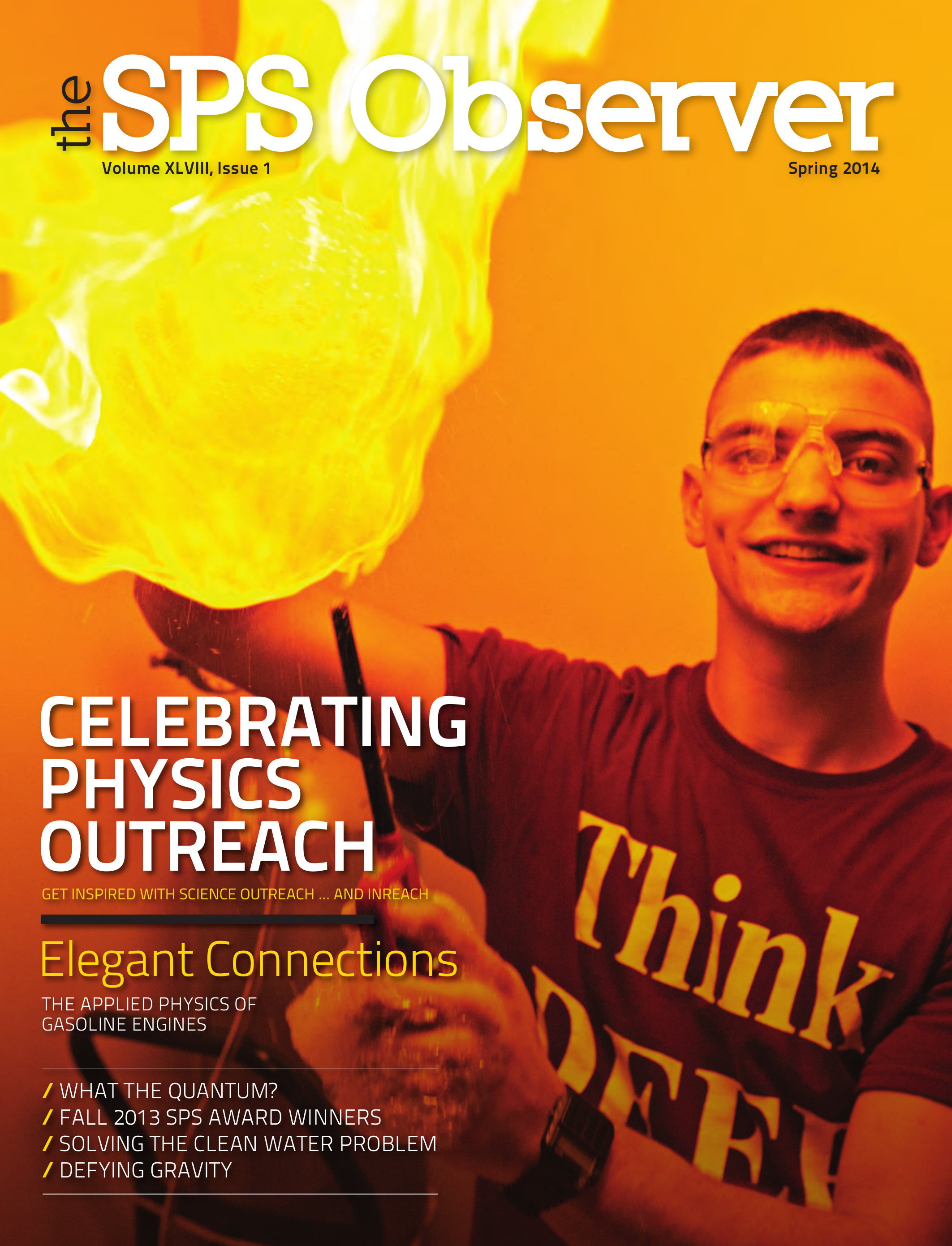}
\caption{\label{bubbles} The front cover of the SPS
Observer features the DEEP student showing his demonstration at Texas A\&M Physics \& Engineering Festival \cite{SPS2014}. Reproduced with permission from the SPS. Photo by Igor Kraguljac.}
\end{figure}

Now that we've discussed the DEEP program in depth, we will briefly review four other programs which share some of the same program principles as DEEP. All of these programs work together synergistically to build a cohesive set of year-long outreach opportunities for students. 

\textbf{\emph{The Texas A\&M Physics and Engineering Festival}} \cite{Festival} founded in 2003 is an annual event that attracts over six thousand visitors yearly. K-12 students and their families from all over Texas and nationwide attend the Festival; many schools bring busloads of students. For schools with a large percentage of underrepresented minority students transportation is partially paid through university diversity grants. The Festival includes a weekend on campus packed with activities: hundreds of hands-on demonstrations, juggling science circus, bubble shows, meetings with astronauts, and public lectures by world renowned physicists. Previous speakers included Stephen Hawking (twice), Brian Greene, Phil Plait, Sean Carroll, Lucianne Walkowicz, Robert Kirshner, Rocky Kolb, Dudley Herschbach, and many others. Visitors appreciate the opportunity to tour the Texas A\&M Cyclotron Institute, interact with Nobel Laureate David Lee in his research lab, and (virtually) tour the Large Hadron Collider. The Festival is a member of the Science Festivals Alliance, a collaboration of institutions committed to serve the public through informal science venues \cite{SFA}.

Hands-on demonstrations run by DEEP students and other student volunteers are the heart of the Festival and the primary reason why people attend the event. Several hundred undergraduate and graduate student volunteers participate in the Festival, explaining physics concepts behind interactive hands-on demonstrations for seven hours. The Festival gives students an opportunity to explain physics concepts to children and adults. The Festival dissolves the boundaries between different populaces in academia: whether you're a freshman in your first physics course or you're a Nobel Laureate, everyone works together as a team at the festival, building excitement for science and technology with the crowds who show up. All these contexts provide an opportunity for transformational experiences.  

\textbf{\emph{Texas A\&M Physics Show}} \cite{Show} is another venue for students to present interactive hands-on demonstrations. The Physics Show (2007 - current) is an entertaining and educational presentation adjustable to any audience level. There are two parts: one-hour presentation followed by 30-minute interactive hands-on activities (mini Festival). Two physics majors help with the presentation and 5-10 graduate and upper-level undergraduate students lead the hands-on part. There are an average of 40 Physics Shows per year attended by 3,000 K-12 students. 

\textbf{\emph{Just Add Science and Game Day Physics}} \cite{AddSciTAMU,Game} are outreach programs in which the students “meet people where they are” \cite{SFA}, by bringing their favorite hands-on demonstrations to existing events and venues where people are already gathered: home football games, heritage and community festivals, etc. These efforts engage with audience members who may never attend a science event on their own accord. The students work as a well-coordinated team and explain physics concepts to every interested person who passes by.

In the \textbf{\emph{Real Physics Live}} program \cite{Videos} students create short entertaining videos about physics demonstrations explaining the underlying physical principles. The videos are intended for middle and high school students, college freshmen, the general public, and all physics enthusiasts. Graduate and undergraduate students work as a team to write scenarios and then star in the videos. 

All programs have similar principles: through participation in these programs the students design and build, teach/serve the public by applying their physics knowledge, communicate scientific principles to non-scientists in an exciting way, lead, work in teams, and last but not least, have a chance to build connections across academic levels (undergraduate-graduate-postdoc-faculty). 

It should be noted that one of the authors is the founder and organizer of several programs described in this section. Two other authors are former active participants in multiple programs. The DEEP program was partially supported by a Tier 1 grant from Texas A\&M and by the Texas A\&M University system. Real Physics Live was supported by a mini outreach grant from APS. Just Add Science was partially supported by a grant from the Science Festival Alliance. All programs received ongoing support from the Department of Physics \& Astronomy at Texas A\&M. 

\section{Framework}



Since our goal is to identify the effects of facilitating informal physics programs on the development of university students, we needed three theoretical frameworks: one to define learning, one to define identity, and one to explain how powerful learning happens. These frameworks were important in developing the instruments used to analyze student development within the programs described in Sec. II. These instruments are described in further detail in Sec. IV. 


We define learning through the framework of situated learning theory in which learning is defined as increased patterns of participation in a community of practice and identification as a member of the community \cite{LaveWenger1991}. One of the key insights of situated learning theory is that learning begins through legitimate peripheral participation. Newcomers to a community observe the community from the periphery and gradually participate more as existing members of the community mentor them into community ideas and practices. Over time they develop greater proficiency in knowledge, using new ways of knowing, and adopting the practices of the community as they move from the periphery toward the core \cite{Wenger2000}. This movement is characterized not only by greater expertise, but more importantly by being recognized (by themselves and others) as members of the community. As their position approaches the core of the community, they take on role identities as leaders and mentors with greater visibility and responsibility \cite{Wenger2015}. While situated learning theory defines the nature of learning, we need an additional framework to understand how learning takes place. 


Transformative learning theory provides a useful lens through which to understand how powerful learning occurs. This theory describes powerful learning as a process of the transformation of one’s “frames of reference” \cite{Mezirow2009}. These frames of reference include assumptions, beliefs, perspectives, mindsets, and habits of mind \cite{Kegan2009}. Through situated learning theory and transformative learning theory, we see learning as a process of becoming, engagement in the practices and knowledge of a community, and a change in identity. 

We define physics identity through the lens of a Dynamic Systems Model of Role Identity (DSMRI) which embraces the complexity of social-contextual elements which interact to facilitate or inhibit identity change \cite{Kaplan2017}. In Hazari et al \cite{Hazari2010}, students' physics identity was defined as belief in ability to understand physics content, recognition by others as being a good physics student, and interest as demonstrated by desire/curiosity to think about and understand physics. We use this definition as a seed and expand it through the framework of DSMRI which characterizes identity as context-specific self-perceptions, values, beliefs, goals, emotions, and perceived action potentials \cite{Kaplan2017}. Facilitation of identity development in learning environments requires contextual features for triggering identity exploration, scaffolding identity exploration, promoting relevance, and facilitating a sense of safety \cite{Kaplan2014}.  

Through the lens of the three frameworks described above, we define physics learning as a process of change characterized by changes in engagement in the physics community, increasing identification as a member of the physics community, transformation of perspectives about the nature and role of physics in society, and physics role identity development.

\section{Methods}
The frameworks described above were used to design a mixed methods study to investigate the impact of Texas A\&M's informal physics programs on student facilitators. A survey was developed to investigate the impact on student's identity, sense of belonging to the physics community, and 21$^{st}$ century skills. This survey consisted of a subset of items from the Colorado Learning Attitudes about Science Survey (CLASS) as well as additional items constructed for the broader goals of this work \cite{adams2006new}. The survey was distributed via email to current and former students who had worked with at least one physics outreach program between 2013-2019. Two follow-up emails were sent at two week intervals after the initial survey to encourage as many responses as possible. At the end of the survey, respondents were asked if they were willing to be contacted for a follow-up interview. Survey responses were analyzed for descriptive statistics, specifically looking at self-reported connections between experiences facilitating outreach and improvements to both physics and non-physics abilities.

The survey was distributed to nearly 400 current and former students. A total of 117 completed survey responses were received. As seen in Table \ref{SurveyDemo}, just over a quarter of responses were from female students, with 1\% identifying as non-binary. A majority of responses, 62\%, were from current or former undergraduate students and 43\% were from current or former graduate students. The excess percentages of responses were due to students who completed their undergraduate work at Texas A\&M and were either still completing or had also completed their graduate work there.   

Interviews were conducted with 35 current and former students recruited from the volunteer pool of respondents from the survey. This number of interviewees is more than adequate for this study since 97 percent of themes in interview-based case studies are identified after twelve interviews \cite{Guest2006}. Interview questions were created based on the three frameworks described in Sec. III that provide a definition and explanation of learning, permitting a multifaceted approach to investigating identity. These questions probed for more in-depth student experiences during facilitation of outreach programs. Interviews were conducted remotely and typically lasted 15-30 minutes. Interviews were conducted by a researcher who was unfamiliar with each interviewee. 

Interviews were coded using a code book based on the frameworks described in Section III. A total of 64 codes were used to categorize statements by three members of the research team in MAXQDA. These codes were organized into categories of i) community (e.g. connecting with participants, accountability), ii) soft skills (e.g. creativity and innovation, teamwork and leadership, communication), iii) hard skills (disciplinary and non-disciplinary skills), iv) affect and experience (e.g. seeing new perspectives, transformational experiences, motivation and excitement, and v) identity (e.g. curiosity, positionality related to ethnicity and gender, worldview, becoming more confident). Categories i-iii are grounded in situated learning theory, category iv is grounded in transformational learning theory, and category v is grounded in DSMRI. 

Coding of interviews was done in stages by different members of the research team. Initially, a team of six coded three interviews after which the coding process was discussed and revisions were made to the code book. A group of three researchers then coded the same five interviews. Comparisons between the three researchers yielded reliability kappa values of $>$0.8 between the first two researchers and $>$0.6 between each of the first two researchers and the third. One researcher coded all thirty-five interviews, while the remaining two researchers each coded a subset of interviews. All interviews were coded by at least two researchers.

To explore the relationships between codes we employed semantic network analysis. Semantic network analysis uses social network analysis tools, but instead of analyzing interactions within networks of people, it focuses on interactions within networks of ideas \cite{DonaldsonAllenHandy2019}. Analysis of these networks starts with determining the likelihood that one idea (a coded segment of text in the data) appears in the text near another idea (a different coded segment). Pearson’s correlations of code co-occurrences were calculated for each pair of codes, producing a correlation matrix at the confidence level p$<$0.01 and another correlation matrix at the confidence level p$<$0.001. These correlation matrices were imported into the UCINET network analysis software \cite{UCINET2002} as 1-mode networks, in which the columns and rows are identical and cells indicate the ties between each set of codes \cite{Scott2017}. In order to characterize the position of any given idea within the network of ideas, the centrality of each idea in relation to other ideas in terms of distance (steps) between that idea and all other ideas was measured by using UCINET to produce Eigenvector centrality measures \cite{Kadushin2012}. These measures were then visualized in NetDraw \cite{NETDRAW2002} as semantic network maps in which each idea (node) is positioned in relation to direct ties — other ideas which are significantly correlated — as well as indirect ties in the form of ties with other ideas which are not significantly correlated directly with the given node, but have ties with the other node through intermediary nodes which are significantly correlated with both \cite{Valente2010}. For instance, if A is correlated with B, and B is correlated with C (but C is not correlated with A), A has an indirect tie with C. This is equivalent to a node level analysis used in prior physics education studies \cite{Brewe2018}. These semantic network maps allow the researcher to understand the complex relationships between ideas, as well as characterize the importance of ideas within the network by automating node size as a function of Eigenvector centrality measures. Additional analysis can help the researcher identify groups of interconnected ideas. Girvan-Newman cluster analysis \cite{Newman2004} in NetDraw uses betweenness centrality measures of links to visually identify clusters (color coding), thus allowing the researcher to describe semantic themes and relationships between themes.

\section{Results}
\begin{figure}[h]
\centering
\includegraphics[width=8.6cm]{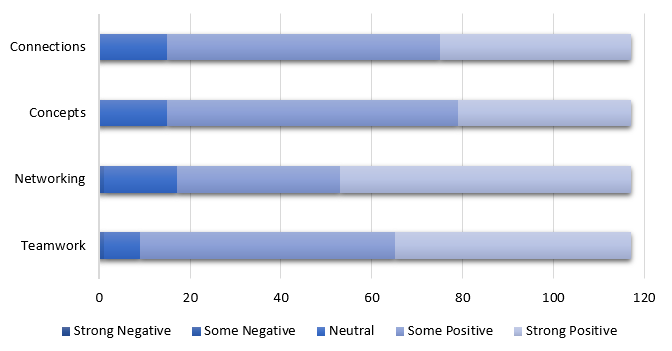}
\caption{\label{Survey} Survey responses from all students on the impact of participating in outreach programs on recognizing connections between topics, their depth of conceptual physics understanding, networking within the department, and development of teamwork skills. }
\end{figure}

Students self-reported on the impact of their participation in informal physics programs (which we will refer to as outreach for the rest of the paper) on their depth of understanding, connections between topics, and the development of networking and teamwork skills. Responses to these survey items are shown in Figure \ref{Survey}. A large number of responses indicate that facilitating outreach events had either some positive impact or a strong positive impact on the dimensions listed above. Over 80\% of students reported some positive or a strong positive impact on recognizing connections between physics topics and on their overall understanding in physics. A slightly higher percentage, 85\%, reported that participating in outreach had a positive impact on their teamwork skills and ability to network within the department.

Students were asked to rate their confidence in their choice of majoring in physics before and after participating in outreach. Of the 62 physics students who responded, 29 students indicated an increase in confidence in choosing physics as a major after participating in outreach. Thirty-two students maintained the same level of confidence, ranging from not confident at all (1) to slightly confident (5) to moderately confident (11) to extremely confident (15). One response indicated a decrease in confidence in choice of major.

Students who volunteered to be interviewed were able to elaborate on their experiences in outreach and the connections noted in the results above. The demographics of interviewees are shown in Table \ref{SurveyDemo}. Below we present results on the frequencies of certain codes and themes from the interviews, a semantic network analysis, and meaningful student quotes related to our framework.

\begin{figure}[h]
\centering
\includegraphics[width=8.6cm]{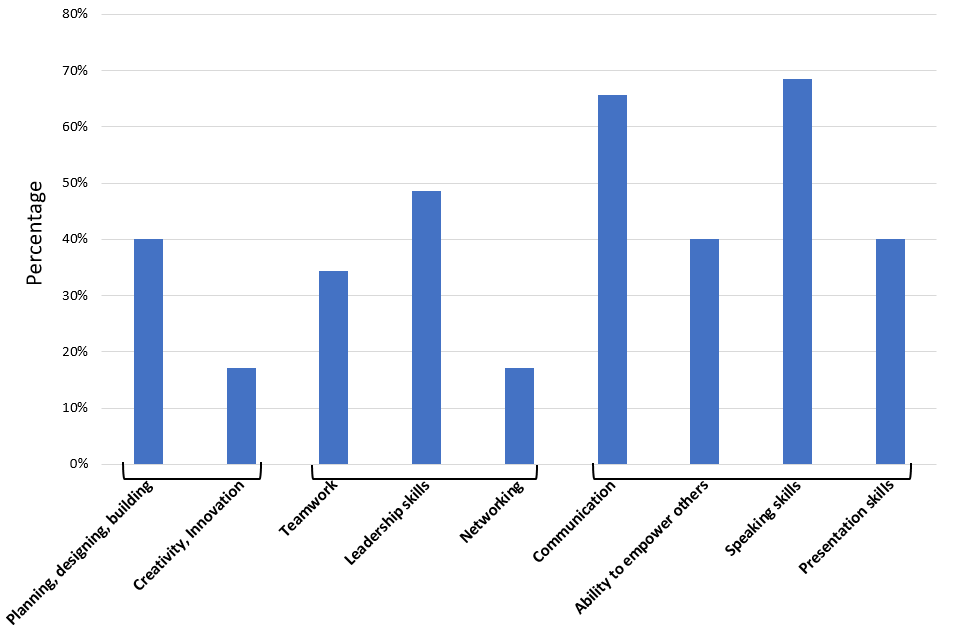}
\caption{\label{Frequency_codes} Frequency of soft (transdisciplinary) skill codes for all interview participants.}
\end{figure}

Interview questions posed to participants probed connections between participation in outreach and their physics identity, values, perceptions, and abilities. Students frequently discussed the impact of outreach on the development of skills related to communication, teamwork, and design. The frequency of codes associated with these skills from all interviews are shown in Figure \ref{Frequency_codes}. A significant number of interviewees touched on the impact of outreach on their communication skills, particularly their speaking ability. From the experience of one graduate student,
\begin{quote}
    \textit{I've really learned over time that it's one thing to know something, but it's a whole different thing to be able to explain it to somebody and really effectively communicate your ideas.}
\end{quote}

Our findings suggest that the student experience in outreach promoted communication not only with other people in STEM, but also with diverse audiences from the general public. From the student perspective, these interactions with the public could provide high impact experiences. Multiple interview participants noted that presenting scientific concepts to such a wide range of people played a significant role in their ability to effectively communicate these challenging concepts with others. As an undergraduate student said,
\begin{quote}
    \textit{If you're going to tell something to a 5 year-old and then something to a 65 year-old right beside them, they both have to understand and they both want something different. You have to learn how to speak on their level and sort of give your audience what they need.} 
\end{quote}

As seen in Figure \ref{Frequency_codes}, communication is not the only skill commonly discussed by students. Leadership and teamwork experience were mentioned by nearly 50\% and 35\% of interviews respectively. In learning to become part of a collaborative effort, one student shared that they \textit{``learned that you can't do it all yourself...that you have to lean on others and be part of a team."} The programs in which these students engaged, such as DEEP mentioned in Section II, can provide new experiences in which to develop interpersonal skills not often found in the classroom. As another graduate student put it,
\begin{quote}
    \textit{I've done teaching, I've done outreach, itself, but … managing and delegating was something that I was not too familiar with, and it definitely gave me very valuable experience ... will be very useful as I continue in my PhD.}
\end{quote}
Establishing interpersonal connections also goes beyond small teams, to networking with the broader department. Students, particularly undergraduates, get a chance to develop additional, and potentially deeper relationships with researchers and faculty through outreach. One undergraduate's experience was that they \textit{``developed a very close working relationship with certain professors in the physics department as a result of [outreach]."}

Though mentioned less frequently during interviews, skills related to creativity and design represent important experiences. Students engaged in building new, or improving existing, demonstrations must develop and implement new solutions to each project. These projects often build on skills developed first within a current or previous course. From one student's experience,
\begin{quote}
    \textit{You build demonstrations so you have to have a plan for them, put together an electrical schematic in order to have an Arduino-powered thing. This is all very real-world application stuff, and it all works really good on resumes.}
\end{quote}

\begin{figure*}[h]
\centering
\includegraphics[width=17.0cm]{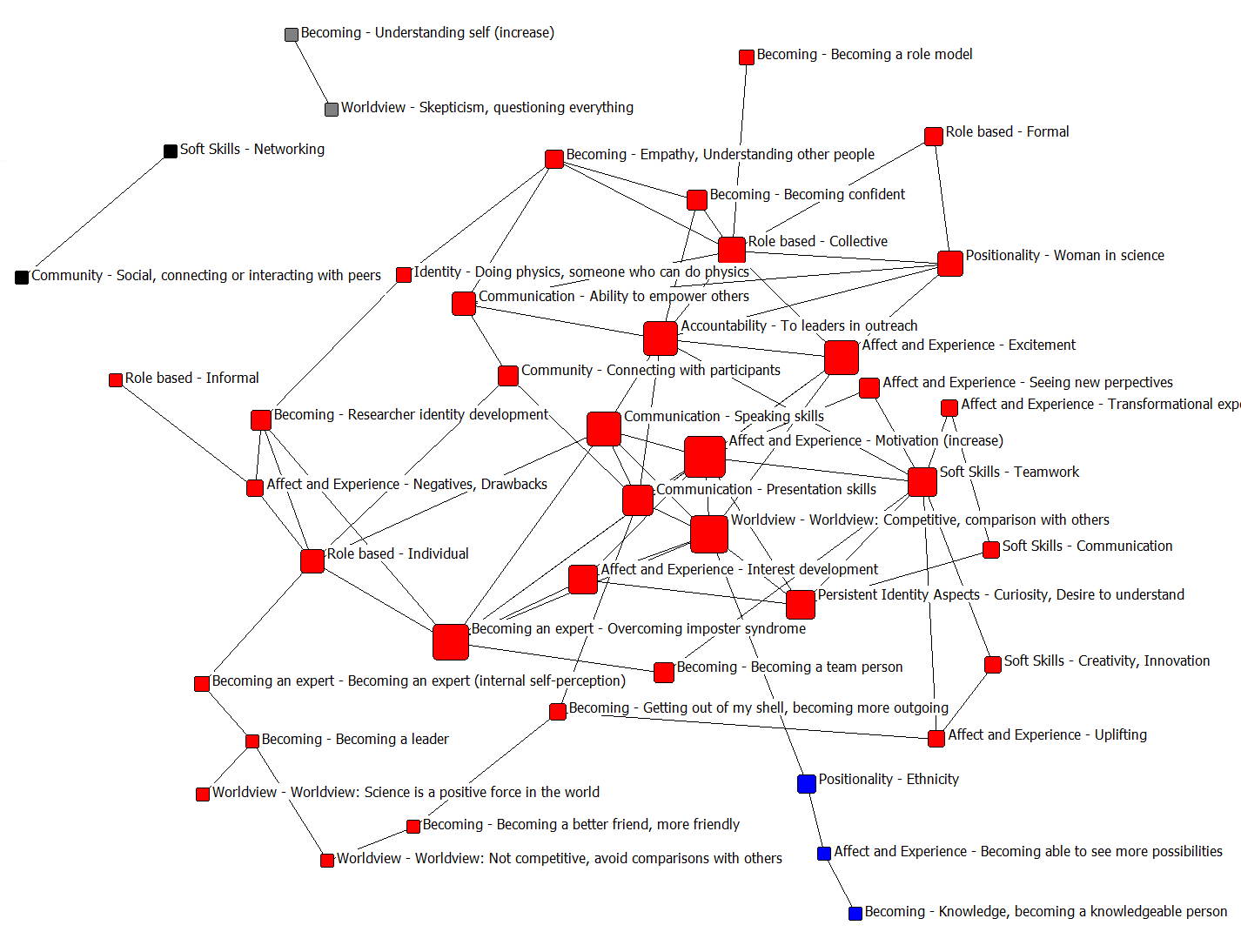}
\caption{\label{Network_map} Semantic network map comprising all codes from 35 interviews at the level of p $<$ 0.01.}
\end{figure*}

To identify emergent themes and significant links between codes, a Girvan-Newman cluster analysis was performed. The resulting semantic network map, at the p$<$0.01 level, is shown in Figure \ref{Network_map}. Here the interconnectedness of major themes is evident, with larger blocks and a higher number of links representing centrality of a code to the impact of students' experiences due to outreach.

A number of important themes and connections are observed from the map in Figure \ref{Network_map}. The most central node is increased motivation, which is linked to nodes related to both physics identity and skill development. Motivation will drive student engagement in outreach, providing opportunities to impact their physics identity and skill development. 


During interviews, students frequently mentioned the growth of their ability to explain and present topics to a variety of audiences, which ranged from young children to adults and sometimes included physics faculty and researchers. Students also touched on their growth in comfort and ability to work as part of a team. Adjacent to motivation, teamwork acts as a further nexus between several important themes of curiosity, ability to see new perspectives, creativity and innovation, as well as the potential for students to have a transformational experience. As an example, we consider the reflections of a graduate student who said, 
\begin{quote}
    \textit{but what I learned through years of doing outreach is that instilling a sense of awe and fascination in entire classrooms full of kids is way, way more important than coming up with some new physics law.}
\end{quote}

Another major theme was students coming to view themselves as more of an expert in physics. Interacting with others could help students see themselves as a physics person because \textit{``other people saw"} them \textit{``as a physics person"}. This node shares several important links to communication skills, which could help develop a sense of expertise, as well as aspects of identity related to becoming a team person and the development of an identity as a researcher. This link to researcher identity would be particularly impactful for undergraduate students considering a graduate degree or current graduate students engaged in research. From one student's experience, 
\begin{quote}
    \textit{I literally just once in a while went out and did a demo. Explained it again and again, the whole day. And that was really fun. And so that helped... solidify my image of myself as a physicist.}
\end{quote}

The role of strong leadership through accountability forms an important cluster linked to student confidence and excitement as well as the ability to empower others. Such leadership provides not only the structure for outreach programs, but also has a significant impact on students by being an exemplar for skills and fostering the culture around outreach. One student discussed this impact, stating
\begin{quote}
    \textit{there is a genuinely amazing community at Texas A\&M, and so much of it does center around [Dr. X].}
\end{quote}

Several secondary themes are also evident from the semantic network map. These themes exhibit fewer links and less centrality, but still offer important insights into the impact of participation in outreach. The personal perception of becoming more of an expert in physics is linked to skill development, individual responsibility, and an identity as a researcher. Experiences gained through teamwork show links to the ability to see new perspectives, the potential for transformational experiences, and sense of belonging to the physics community. These are highly valuable aspects of a student's learning. From one undergraduate student's perspective,
\begin{quote}\textit{I want to say that through these outreach [activities], I probably have felt closer to the physics community than I have through my classes themselves.} 
\end{quote}
While another undergraduate student stated their experience as,
\begin{quote}\textit{I think that's affected my identity as a physics person the most. Where I just kind of feel like I'm a part of this community in a sense. Like physics is something that I want to do and engage in.} 
\end{quote}

Two peripheral themes are noteworthy for their contributions to the sense of community among students. Learning to understand others, or become more empathetic, is linked to themes including student confidence, identity as someone who can do physics, and communication. Students who facilitated outreach programs described how it \textit{``impacted [their] ability to connect with others"} in a positive way. Outreach also provided a social environment for students to develop connections with their peers. From the experience of one student
\begin{quote}
    \textit{I went from a more loner type person to being very outgoing and social within the physics community and being able to bond with other people through [outreach] events.}
\end{quote}
For other students, the excitement and demands of outreach events provided a bonding experience with their peers. In the words of one student
\begin{quote}
    \textit{It helped me make a lot of friends...You fight in the trenches with a lot of people. You have these exhausting all day things where you talked to so many people.}
\end{quote}
These experiences, for some students, led to a deeper sense of ownership and connection with physics, helping them want to become better ambassadors for their field. From one student's perspective
\begin{quote}
    \textit{I think it kind of shapes it to where almost being in physics is a fun thing, and it's making me more want to be a representative for the major in a sense.}
    \end{quote}

Being a woman in physics was observed to have strong ties to outreach leadership and seeing oneself as a member of the scientific community. For context with this theme, it should be noted that the coordinator for most outreach programs at Texas A\&M is a female faculty member. In the words of one female undergraduate student, this impact was described thus:
\begin{quote}
    \textit{I think [physics outreach activities] have really made me feel like I can be a part of the physics major. I know as a freshman I felt like maybe this wasn't the right major, anything like this, but I think going out and teaching other people physics made me feel like I knew what I was doing and made me feel like I could keep going on the route of being a physics major.}
\end{quote}

It is apparent that the community created through outreach activities promotes the building of relationships and sense of belonging to the physics community for all students. Although the framework employed in this work did not seek to specifically differentiate the experiences of different groups of students, certain patterns emerged when comparing male and female students. Our analysis suggests that female students experience stronger benefits from interactions with their peers and faculty as well as recognition that there are people like them within the physics community. This feeling of representation, in particular, appears to be linked with a deeper sense of belonging, which can be a critical factor in the determination of student retention in higher education \cite{Kalender2019}. As one undergraduate female student put it:
\begin{quote}
    \textit{I will say that I met a lot of friends through physics outreach. And a lot of them were girls in physics. And it was kind of cool to meet a lot of people who were having the same thoughts as me, and we could just kind of band together and have our own little community within the physics department.}
\end{quote}

A second theme that emerged among female interviewees was the importance of external recognition on physics identity, which may also relate to physics self-efficacy. In the words of one female graduate student \textit{``It helped me see myself as a physics person because other people saw me as a physics person."} Whereas male students spoke more often and more directly about internal self-perception, or viewing themselves as experts in the field, female students self-perception was discussed more in the context of recognition as an expert from others. Outreach provides the opportunity for all students to display their expertise to public audiences. While there are benefits of this recognition as an expert to all students, the network analysis shows this was more impactful to female students. As one undergraduate female student put it,
\begin{quote}
    \textit{So it's like not only do I believe in myself, but I have others who believe in me, so that way if I ever falter in my belief in myself, I can fall back on the other people who believe in me.}
\end{quote}
The impact of recognition on students who participated in outreach was important, but was of a different nature for female students than for male students.

\begin{table}[h]
\caption{\label{SurveyDemo} Demographics of survey responses by student gender and classification. A total of 117 complete survey responses were received and 35 interviews were completed. Numbers add up to more than 117 as a few students continued as graduate students at Texas A\&M after completing their undergraduate degrees.}
\begin{ruledtabular}
\begin{tabular}{ccccc}
  & \multicolumn{2}{c}{Survey} & \multicolumn{2}{c}{Interview} \\
  & Male & Female & Male & Female \\
 \hline
 Current Undergraduate & 29 & 17 & 3 & 5\\
 Former Undergraduate & 20 & 4 & 3 & 1\\
 Current Graduate & 21 & 6 & 11 & 3\\
 Former Graduate & 14 & 5 & 8 & 2\\
\end{tabular}
\end{ruledtabular}
\end{table}

\section{Discussion}
The informal physics programs described in Section II share common traits of being student-focused programs that promote peer interactions, sense of community, individual growth, and service to the public. It must be noted that most of these programs  historically started in response to the requests from the community and local schools. Then it became clear that they could be sustainable only with a maximally broad involvement of student population. Therefore, a concerted effort has been made to convert their involvement into a meaningful learning experience by fostering the principles described in Figure \ref{Programs}. The Texas A\&M physics outreach programs focus on students designing, building, and presenting demonstrations to audiences of all ages and in a variety of settings. The largest annual event includes participation from several hundred STEM students and has welcomed around 7,000 attendees in recent years. Smaller events throughout the year involve 2-20 physics students engaging with groups ranging from K-12 students making visits to campus to adults of diverse backgrounds at local community events and festivals. 

The results noted in the prior section suggest that students facilitating outreach programs experience positive impacts on their individual physics identity, enhance their understanding of the concepts of the field, increase their confidence, sense of belonging to the physics community, and improve their 21st century skills. Our results showed that many students who facilitated physics outreach programs reported an increase in their confidence with physics as their choice of major. Through working with diverse audiences, students developed a sense of their own expertise within physics leading to the development of their personal physics identity. 

We found that student motivation was a strong central theme. As follows from Fig. \ref{Network_map}, this motivation is directly connected with strong central nodes of student excitement, curiosity, and changes in their worldview; development of interest in physics, communication and teamwork skills; becoming an expert. These are all building blocks of situated learning, transformative learning, and physics role identity as defined in Sec. III. One can conclude that the students who facilitated informal programs experienced a positive impact on their physics identity. This parallels recent results from Fracchiolla et al. who reported a connection between volunteering with physics outreach programs and the development of physics identity \cite{Fracchiolla2020}.



Accountability to leaders in outreach is another strong central theme in Fig. \ref{Network_map}. Analysis of interviews showed that the presence of a strong role model (the faculty member in charge of outreach programs) was central in promoting excitement among students and increased their motivation to engage in outreach events. 


Analysis of interviews suggested that external recognition is particularly impactful for female students, facilitating growth of their individual physics identities. This external recognition came from both the community of practice surrounding informal physics programs as well as from the audience that students were interacting with. In contrast to previous work by Hyater-Adams et al., interviews with female students in this study only reported positively on the effects of recognition from others \cite{Hyater-Adams2018}. These interactions reinforced their identity as a physics person. This could indicate that outreach is a high reward engagement for retention of female students since, as noted by Hazari et al., the development of a physics identity can help students choose and persist within the field of physics \cite{Hazari2010}. 

From the recent JTUPP report, a number of skills were identified as being high priority for preparing students for a 21st century workplace, including communication and teamwork \cite{Heron2016}. During interviews, students frequently discussed how facilitating outreach had improved their ability to communicate with others. This is an essential part of a student's development, as any scientist should be proficient in communicating to a diverse set of audiences. Whether a physicist works in academia, industry, government, or elsewhere, it is an essential skill to effectively share ideas with others, whether they are knowledgeable about our field or not. In having students engage in outreach they become teachers to their audiences. This provides not only an essential skill for the 21st century, but can also be a critical reinforcement of the formal physics training of a degree plan \cite{Feldon2011}. As noted by Hinko and Finkelstein, outreach is a generally overlooked area for the development of teaching skills \cite{Hinko2016}. Outreach provides the structure through which we can let students teach. 

Many outreach programs offer opportunities to foster teamwork and leadership skills on an ongoing basis. While small group projects are often incorporated into formal courses - such as labs - for short-term projects, much of a physics curriculum focuses primarily on the work of the individual. This is true at the undergraduate level, and especially true at the graduate level. Working on a demonstration, or having ownership of a group of demonstrations, is a task that goes beyond a single course and unit. Teams often work for most of an academic year to research, design, and build the first version of a demonstration. For many projects, there is often a second (and sometimes third) cycle to improve a demonstration. Our findings show that outreach supports the development of teamwork and leadership skills within a low stakes environment. There is a particularly strong benefit to graduate students through the DEEP program in working to manage a team of undergraduate students while receiving mentoring from outreach leaders. These teams are effectively a mock research lab with the graduate student functioning as the principle investigator of the project. 

In addition to communication and teamwork, some outreach program structures support the development of creativity and design skills. When building a new demonstration there is a goal in mind (produce the demo!), but the specific form and method by which this is achieved is ambiguous. As seen in analysis of our interviews, students were able to bring their own vision to projects during the development cycle while learning the "real-world" side to the application of different parts of their education, such as circuits. Giving students ownership of a project which will be used long after their time as a student provides a sense of legacy and commitment to the apparatus that they work to create. Such experiences have been recommended as ways to augment undergraduate science education, meeting the broad range of individual interests and talents of students \cite{Thiry2011}. 

Sharing physics with people of all ages can help students develop a better conceptual understanding of physics and recognize more connections between different areas. In discussing the impact of outreach on their physics understanding, students used words like \textit{``solidify"} and \textit{``cemented"} during their interviews. In sharing their knowledge students restate their understanding, and frame their knowledge in a way that it can be understood. This depth of understanding is also strengthened by responding to questions where students may push the limits of their own understanding to come up with a good bridging analogy or explanation \cite{Redish1994}.

The results of this work are in alignment with previous finding from evaluators at the Education Research Center at Texas A\&M who concluded that DEEP is a ``highly successful" program \cite{Stillisano2014}. Program evaluators concluded that significant enthusiasm existed among undergraduate participants with many indicating intentions to return to the program in subsequent years. Students reported their DEEP experiences as increasing understanding of physics and engineering concepts, in addition to problem-solving skills. Mentorship in the DEEP program demonstrated effective leadership as reported by both graduate mentors and undergraduate students. The DEEP program was identified as on target to enhance undergraduate experiences and support students through active learning, service-oriented learning, and teamwork.

\section{Limitations}

This study expands our understanding of the impact of student participation in informal physics programs, however, there are several limitations which should be noted. First, students self-reported on their experiences and perspectives during both survey and interviews. Also, participants in this study come from a single, large, land-grant, four-year public institution with a diverse undergraduate enrollment. Information on student identities such as ethnicity, first-generation status, etc. were not collected. Only the demographic information of gender or classification (undergraduate or graduate) was asked for. The outreach programs included in this study also represent a subsection of the potential program structures which exist in physics departments across the country. A larger follow-up study should collect this information, which would allow for a deeper analysis of the impact of participation in outreach on different student identities. Furthermore, findings from qualitative analysis from interviews are not generalizeable, following principles of social science research. 

\section{Conclusion}
The transition from novice to physicist is a lengthy and complex process that is guided by both formal and informal experiences. For many, this is a journey that begins with a movement towards becoming a physicist, but subsequently turns in other directions. In this work, we observed how participation in informal physics programs can support an individual in becoming a physicist and boost their development through less structured, but critically important, experiential learning. Informal physics programs provide an environment in which students engage in experiential learning through facilitating physics demonstrations and are able to learn through teaching individuals from a diverse set of backgrounds. These experiences can provide rich teaching opportunities for students by bringing physics beyond the pages of a textbook \cite{Hinko2016}, which challenges them to break down concepts, and potentially promotes a deeper understanding.

We have presented findings from a mixed methods study on the impact of five informal physics programs on a large number of undergraduate and graduate students from a large land-grant university. This study, based on self-reported data, revealed that facilitation of physics outreach programs promoted the development of students' physics identity, sense of belonging to the physics community, and the acquisition and improvement of 21st century skills. Physics outreach programs can provide pathways to enhanced confidence through experiential contexts beyond classrooms and laboratories. By facilitating outreach, students foster skills that promote career readiness such as communication, teamwork and networking, and design skills as well as increased conceptual understanding of physics. There is a significant connection between strong program leadership and multiple themes including skill development, confidence, and motivation. Outreach facilitates the development of a sense of belonging to a physics or STEM community, promoting social interactions beyond formal contexts, such as classrooms or research labs. We observed that female students reported that their sense of belonging in the physics community was linked to interactions with others and external recognition.

We recognize that the informal physics programs included in this study represent a subsection of the potential program structures which exist in physics departments across the country. We believe that the design principles on which these programs rest can become a part of any physics outreach program at any institution. Everyone can start an outreach program, and it doesn't even need a significant budget or a change to current curriculum. Many demonstrations are part of common physics laboratories. These can be used to seed these high impact experiential learning environments which promote student growth as a physicist, as a member of their STEM community, and in the skills they will bring to their future careers. From one student's experience through outreach,
\begin{quote}
    \textit{I wasn't sure if [physics] was a good fit for me, but I've definitely been really reaffirmed that it's something that I want to do and something that I can do, something kind of I'm actually able to do.}
\end{quote}

The impacts of facilitating outreach on physics students merits further examination from multiple perspectives. Our findings suggest that students experience benefits from outreach differently based on gender. A subsequent study that examines the impact of outreach of traditionally underrepresented groups in physics for a larger population of students would be informative. Another useful study would be to look at the impacts of programs of different scales, size, and frequency of events throughout the year. A further dimension that merits attention would be the comparison of facilitating outreach on graduate versus undergraduate students for institutions where these students volunteer side by side. 

\begin{acknowledgments}
We would like to thank all students who participated in the survey and interviews. We would also like to thank the many wonderful students, faculty, staff, and donors who make our outreach programs possible.
\end{acknowledgments}

\bibliography{bibliography}

\end{document}